\documentclass{blois}

\def\Journal#1#2#3#4{{#1} {\bf #2}, #3 (#4)}

\def\be{\begin{equation}}
\def\ee{\end{equation}}
\def\bea{\begin{eqnarray}}
\def\eea{\end{eqnarray}}

\newcommand{\Photo}{}
\begin{document}
\vspace*{4cm}
\title{Tailored PDFs for new physics searches}

\author{ Elie Hammou }

\address{Nikhef Theory Group, Science Park 105, 1098 XG Amsterdam, The Netherlands}

\maketitle\abstracts{
Parton Distribution Functions (PDFs) at large Bjorken-$x$ are mostly constrained by high-energy 
measurements, and thus risk to absorb the
energy-growing effects of unaccounted new physics (NP) if it were present in the high-energy tails of hadron-collider
observables used in global PDF fits. If undetected, such contamination biases the resulting PDFs and can potentially hide
the absorbed NP signals from subsequent searches. We illustrate this effect in a Standard Model Effective Field
Theory (SMEFT) risk-assessment scenario affecting high-mass Drell-Yan production at the High-Luminosity LHC, and
compare two strategies to obtain robust PDFs and SMEFT bounds: conservative fits that exclude data above an
energy cut, and simultaneous fits of PDFs and Wilson coefficients performed with the {\tt SIMUnet} tool. Both
approaches are shown to successfully recover the injected new physics that is otherwise absorbed and hidden by
the PDFs.}

\section{Introduction}
\label{sec:intro}

With no clear deviation from the Standard Model (SM) reported so far at the LHC, indirect searches for physics
beyond the SM (BSM), looking for small, broad distortions of SM distributions rather than a localised bump,
have become an increasingly important complement to direct searches. The Standard Model Effective Field Theory
(SMEFT) provides a model-independent language for such searches: integrating out heavy BSM states of mass
$\Lambda$ generates a tower of higher-dimensional operators $\mathcal{O}_i^{(6)}$ built out of SM fields,
\begin{equation}
\mathcal{L}_{\rm SMEFT} = \mathcal{L}_{\rm SM} + \sum_i \frac{c_i}{\Lambda^2}\, \mathcal{O}_i^{(6)} + \dots,
\label{eq:smeft}
\end{equation}
whose Wilson coefficients $c_i$ can be fitted directly from data.

At hadron colliders, predictions for these observables factorise into a hard-scattering cross section,
computable in perturbative QCD and EW theory and to which the SMEFT operators of Eq.~(\ref{eq:smeft})
contribute, convoluted with the non-perturbative parton distribution functions (PDFs) of the proton,
$d\sigma^{pp\to X} = \sum_{i,j} f_i \otimes f_j \otimes d\hat\sigma^{ij\to X}$. Both indirect NP searches and PDF
determinations therefore rely on the same high-energy tails of collider observables, and the large-$x$ region of
the PDFs, precisely where BSM sensitivity is largest, is among the least constrained by data; complementary
information from future facilities such as the Electron-Ion Collider and the Forward Physics Facility is expected
to help disentangle genuine NP effects from large-$x$ PDF uncertainties~\cite{eicfpf}. This raises a
central question for precision BSM phenomenology: how should
PDFs be determined so that they do not bias a subsequent NP search? This contribution summarises the
results of Ref.~\cite{us}, to which we refer for details.

\section{Motivation of the problem}
\label{sec:motivation}

Since a global PDF fit is normally performed under the assumption that the SM is the correct theory, any
energy-growing NP effect present in the data can in principle be reabsorbed into the fitted PDFs, producing a
{\it BSM-biased} PDF set that mimics the BSM signal without visibly worsening the fit quality of the
data~\cite{greljo,hideandseek}. In turn, using such a PDF set as an input to a subsequent SMEFT fit propagates
this bias and can mask the presence of the underlying new physics.

We illustrate this issue with a closure test: {\tt NNPDF4.0}~\cite{nnpdf40} is taken as the {\it true}
underlying PDF set, and synthetic charged-current high-mass Drell-Yan (HMDY) data at the High-Luminosity LHC
(HL-LHC, $\sqrt{s}=14$~TeV) are generated including the effect of a heavy, universally-coupled $W'$ boson,
parametrised in the SMEFT by the oblique coefficient $\hat W = 8\times10^{-5}$ ($M_{W'}=13.8$~TeV). We then
perform a SMEFT-only fit of $\hat{W}$, with the PDFs held fixed, using two different input PDF sets: the {\it
true} PDFs used to generate the data, and a set obtained by fitting the same synthetic data under the (false)
assumption that the SM alone describes it.

\begin{figure}[htb]
\centering
\includegraphics[width=0.49\linewidth]{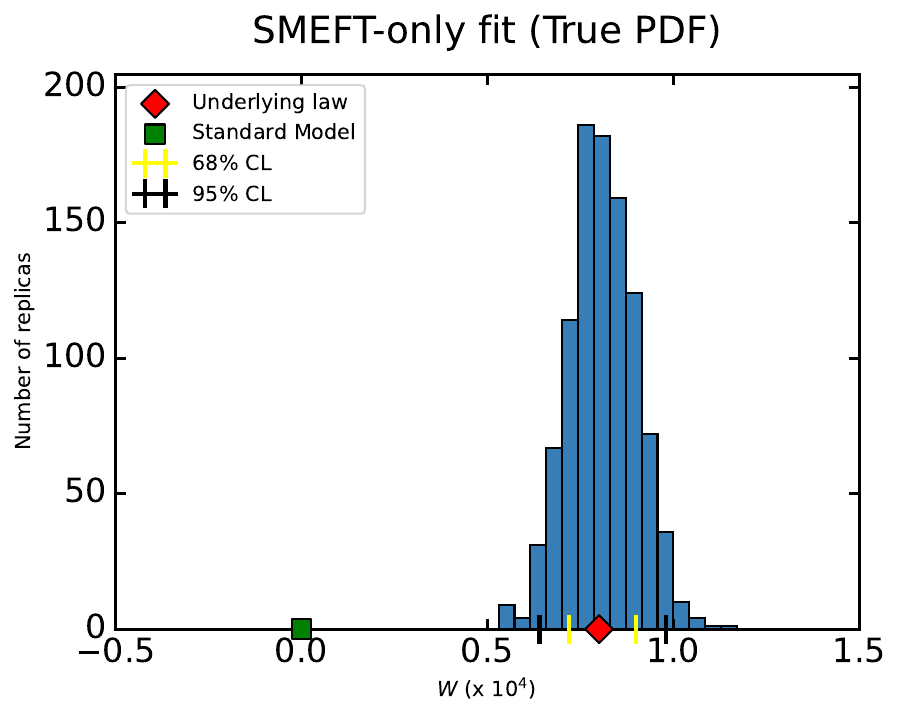}
\includegraphics[width=0.49\linewidth]{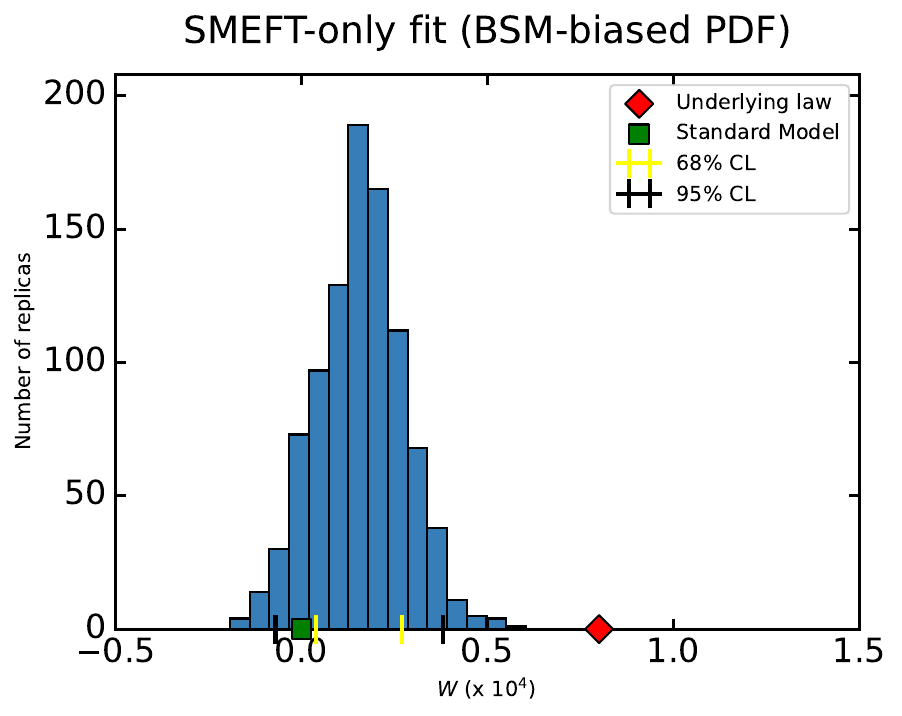}
\caption{Distribution of the fitted $\hat{W}$ coefficient over the fit replicas, using the {\it true} input PDFs
(left) and the {\it BSM-biased} PDFs obtained by fitting the same data under the SM assumption (right). The
injected {\it true} value and the SM point are indicated.}
\label{fig:motivation}
\end{figure}

The results are shown in Fig.~\ref{fig:motivation}. With the true PDFs as input, the fit correctly identifies the
injected $\hat{W}$ and excludes the SM point by more than $5\sigma$. With the BSM-biased PDFs as input, the
picture is reversed: the fit appears compatible with the SM, while excluding the true underlying theory at
similar significance. The signal has been entirely absorbed into the large-$x$ antiquark PDFs, which are only
weakly constrained by pre-HL-LHC data. This shows that a naive, sequential PDF-then-SMEFT fit can hide
new physics rather than identify it, motivating the two mitigation strategies discussed below: {\it conservative}
separate fits, which fit the PDFs and the SMEFT coefficients in two independent steps while explicitly removing
the data that could bias the PDFs, and {\it simultaneous} fits, which determine PDFs and SMEFT coefficients in a
single joint fit, removing the SM assumption.

\section{Conservative separate fits}
\label{sec:conservative}

The most direct fix is to prevent the BSM-sensitive data from entering the PDF fit in the first place. Since the
specific NP model responsible for a distortion is generally unknown, we adopt a model-agnostic strategy in which
all data above a maximum energy scale $Q_{\rm max}$ are excluded from the PDF determination, exploiting the fact
that SMEFT effects generically grow with energy. The resulting {\it
conservative} PDF set is then used as a fixed input to a SMEFT-only fit on the full dataset.

\begin{figure}[htb]
\centering
\includegraphics[width=0.53\linewidth]{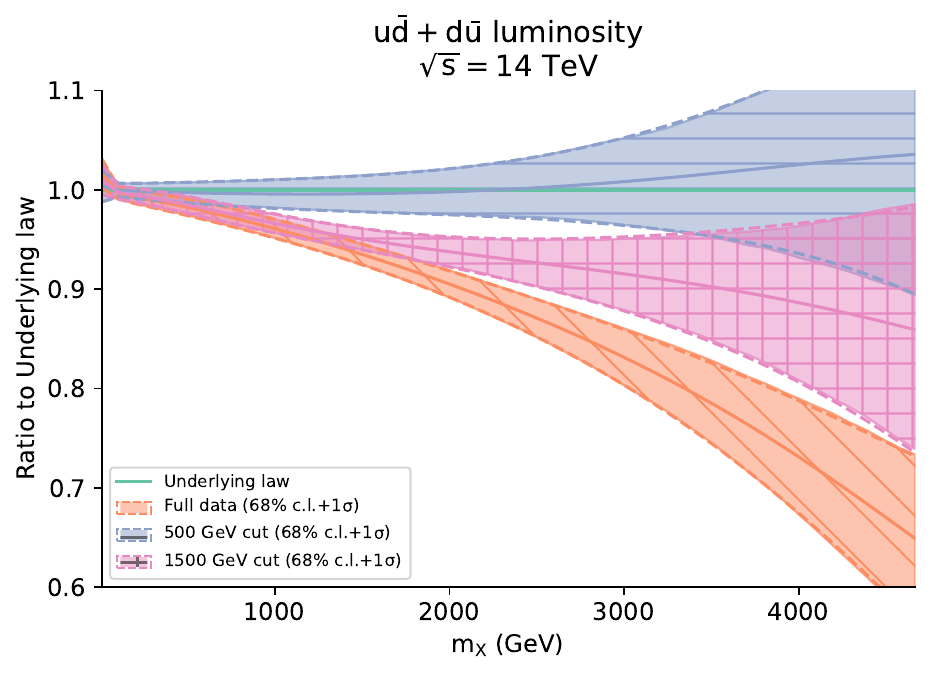}
\includegraphics[width=0.46\linewidth]{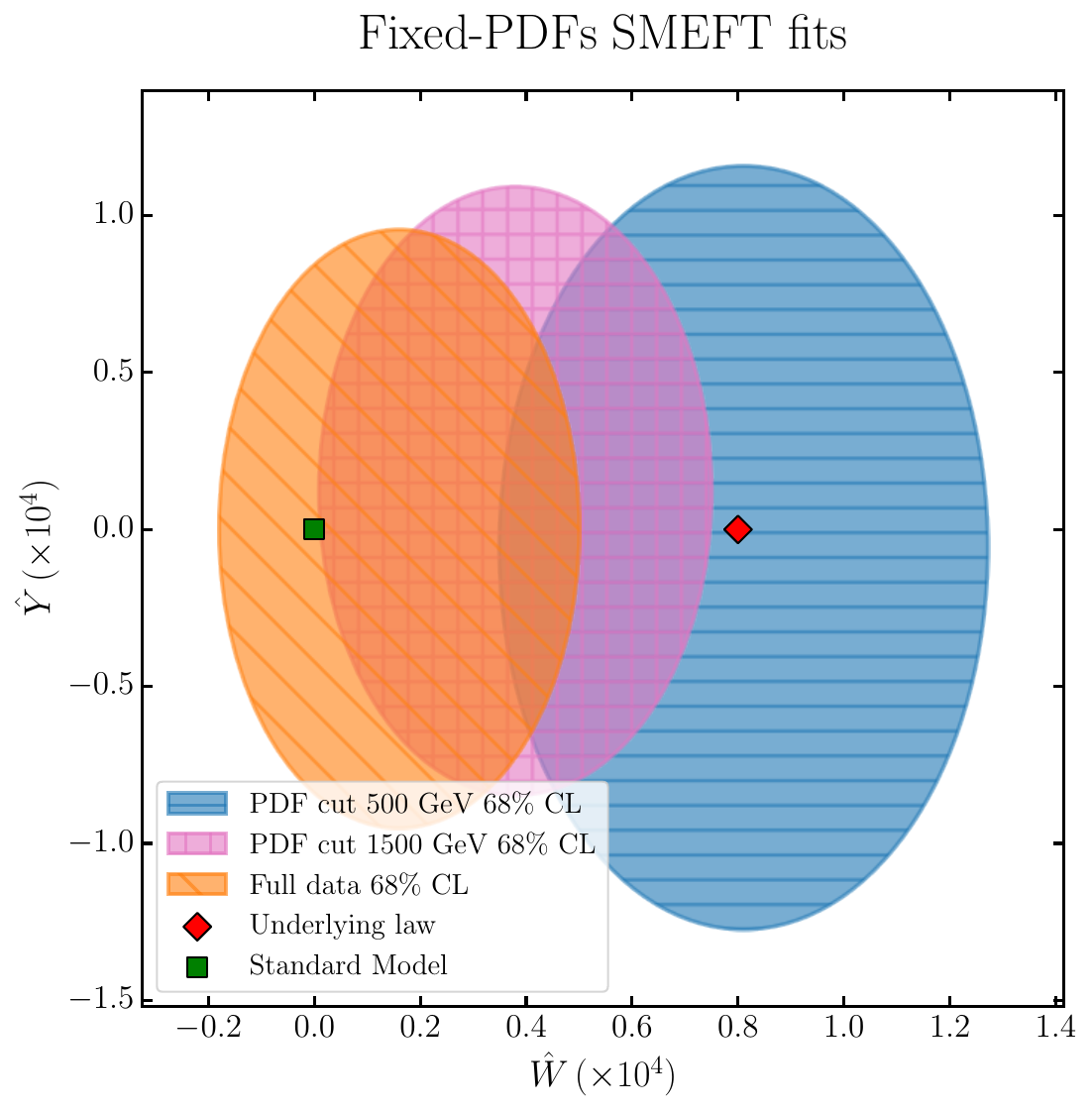}
\caption{Left: $u\bar{d}+d\bar{u}$ PDF luminosity obtained from PDF fits to the $\hat{W}=8\times10^{-5}$
synthetic dataset with no energy cut, and with $Q_{\rm max}=1500$~GeV and $Q_{\rm max}=500$~GeV, compared to the
true underlying law. Right: SMEFT bounds on $\hat{W}$ obtained from SMEFT-only fits using each of these PDF sets
as a fixed input.}
\label{fig:conservative}
\end{figure}

We apply this strategy to the $\hat{W}=8\times10^{-5}$ scenario of Sect.~\ref{sec:motivation}, comparing PDF
fits with no cut, $Q_{\rm max}=1500$~GeV, and $Q_{\rm max}=500$~GeV. As shown in Fig.~\ref{fig:conservative}, a
sufficiently aggressive cut ($Q_{\rm max}=500$~GeV) fully removes the BSM-induced distortion of the PDF
luminosity, and the corresponding SMEFT bound is centred on the true injected value, correctly excluding the SM.
A milder cut ($Q_{\rm max}=1500$~GeV) is not restrictive enough: the bias persists and the fit still favours the
SM over the true theory. As expected, lowering $Q_{\rm max}$ inflates the PDF (and consequently SMEFT)
uncertainties, since less data is available to constrain the large-$x$ region. Varying $Q_{\rm max}$ and checking
the mutual compatibility of the resulting PDFs and SMEFT bounds is itself a useful, model-agnostic diagnostic for
BSM contamination: a systematic, energy-dependent drift as $Q_{\rm max}$ is relaxed is a warning sign, while
purely SM data yields cut-independent results. The main practical drawback of this approach is that there is no
{\it a priori} way of choosing an optimal $Q_{\rm max}$, and once fixed, the high-energy observables that are cut
away can no longer help constrain the PDFs at all.

\section{Simultaneous fits}
\label{sec:simultaneous}

An alternative strategy is to give up the two-step logic altogether and determine the PDFs $\theta$
and the SMEFT coefficients $\mathbf{c}$ concurrently in a single fit, $T(\theta,\mathbf{c}) = {\rm
PDF}(\theta)\otimes\hat\sigma(\mathbf{c})$, using the entire dataset without assuming $\mathbf{c}
= 0$. We use the open-source tool {\tt SIMUnet}~\cite{simunet}, which extends the NNPDF neural-network
PDF-fitting methodology~\cite{iranipour} with an additional SMEFT layer, mapping the fitted PDFs and Wilson
coefficients onto both the SM and SMEFT predictions for each observable within the same fit. Because the fit
no longer assumes the SM anywhere, a BSM signal cannot be selectively absorbed into the PDFs without also
showing up in the fitted Wilson coefficients.

\begin{figure}[htb]
\centering
\includegraphics[width=0.49\linewidth]{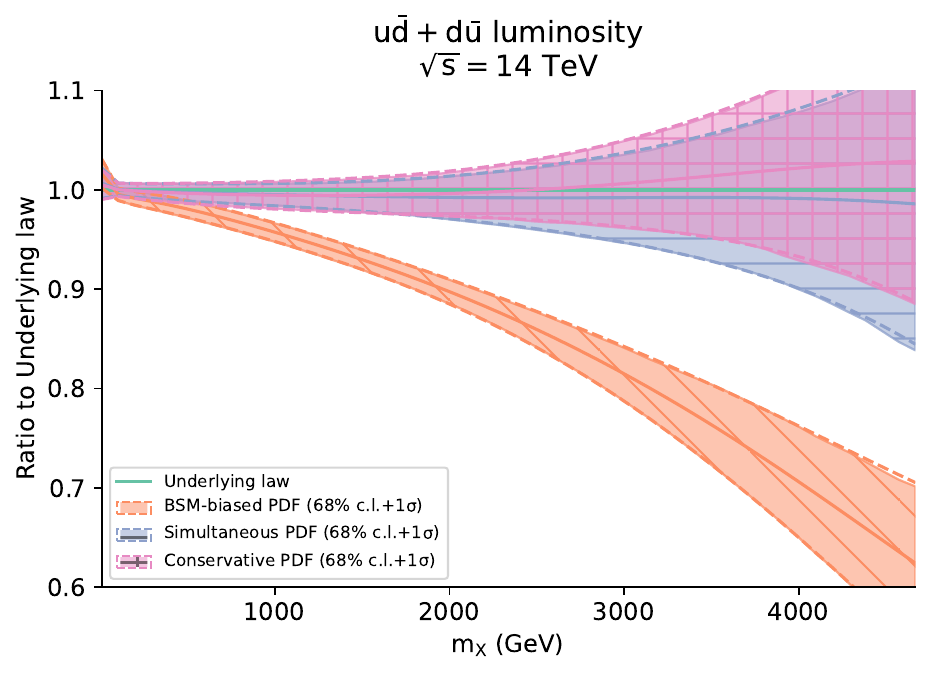}
\includegraphics[width=0.49\linewidth]{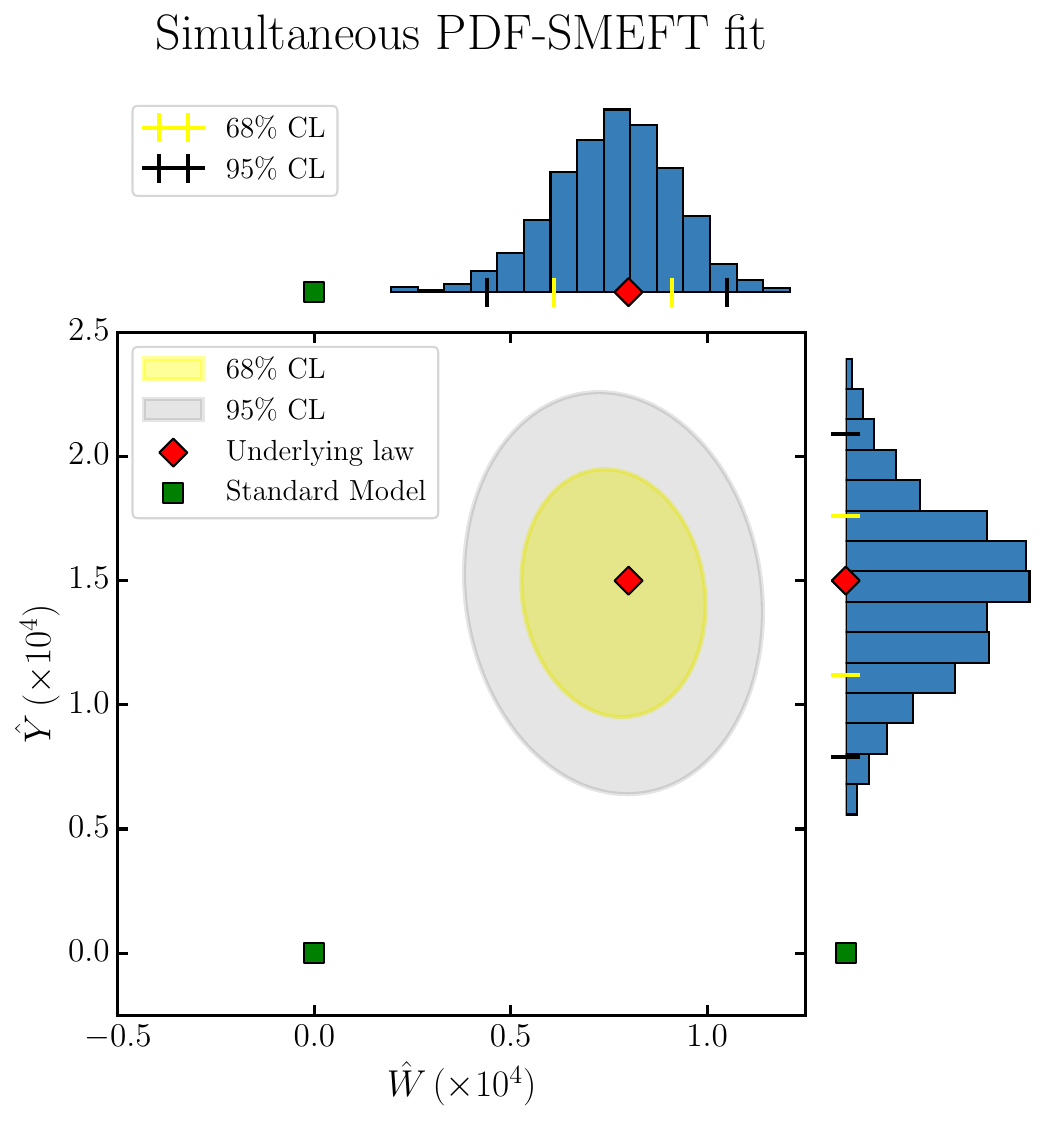}
\caption{Left: $u\bar{d}+d\bar{u}$ PDF luminosity obtained from the BSM-biased fit, the conservative fit
($Q_{\rm max}=500$~GeV) and the simultaneous PDF+SMEFT fit, compared to the true underlying law. Right: 68\% and
95\% C.L. bounds on $(\hat{W},\hat{Y})$ from the simultaneous fit, together with the true and SM points.}
\label{fig:simultaneous}
\end{figure}

We repeat the closure test of Sects.~\ref{sec:motivation}-\ref{sec:conservative}, this time injecting two
correlated SMEFT directions, $\hat{W}=8\times10^{-5}$ and $\hat{Y}=1.5\times10^{-4}$ ($M_{Z'}=18.7$~TeV), into
the full HL-LHC Drell-Yan dataset. Figure~\ref{fig:simultaneous} shows that the simultaneous fit recovers a PDF
luminosity consistent with the true underlying law, with uncertainties comparable to the conservative fit, and
correctly excludes the SM point in the $(\hat{W},\hat{Y})$ plane at high significance. Despite fitting hundreds
of additional PDF parameters together with the Wilson coefficients, the uncertainties on $\hat{W}$ and $\hat{Y}$
remain competitive with the conservative approach, and the correlation between the two SMEFT directions is
reduced, since the full dataset, rather than only the sub-threshold subset, constrains the PDFs. We find
the same qualitative outcome in an independent scenario in which a heavy colour-octet $\hat Z$ affects the
high-invariant-mass tail of $t\bar{t}$ production and is absorbed into the large-$x$ gluon: both the conservative
and the simultaneous fit again recover the injected signal, with the simultaneous fit yielding the tighter bound.

Simultaneous fits thus avoid having to guess an energy cut and make full use of the high-energy tails to
constrain the PDFs, at the price of a higher-dimensional fit and the need to specify in advance which SMEFT
directions to fit for. Both this approach and the conservative fits of Sect.~\ref{sec:conservative} are
effective, complementary strategies for obtaining PDFs and SMEFT bounds that are robust against hidden new
physics; a more complete discussion, together with a $t\bar{t}$-sector case study and further diagnostic
observables, is given in Ref.~\cite{us}.

\section*{Acknowledgments}

I am grateful to my collaborators on Ref.~\cite{us}: E.~Cole, M.~N.~Costantini, L.~Mantani, F.~Merlotti,
M.~Morales-Alvarado and M.~Ubiali. I have been supported for this work by the European Research Council under the European Union's
Horizon 2020 research and innovation Programme (PBSP, Grant agreement n.950246) and partially by the Swiss
National Science Foundation.


\section*{References}

\begin{thebibliography}{99}

\bibitem{us}
E.~Cole, M.~N.~Costantini, E.~Hammou, L.~Mantani, F.~Merlotti, M.~Morales-Alvarado and M.~Ubiali,
arXiv:2602.20235 [hep-ph].

\bibitem{hideandseek}
E.~Hammou, Z.~Kassabov, M.~Madigan, M.~L.~Mangano, L.~Mantani, J.~Moore, M.~Morales Alvarado and M.~Ubiali,
\Journal{JHEP}{11}{090}{2023}, arXiv:2307.10370 [hep-ph].

\bibitem{greljo}
A.~Greljo, S.~Iranipour, Z.~Kassabov, M.~Madigan, J.~Moore, J.~Rojo, M.~Ubiali and C.~Voisey,
\Journal{JHEP}{07}{122}{2021}, arXiv:2104.02723 [hep-ph].

\bibitem{simunet}
M.~N.~Costantini, E.~Hammou, Z.~Kassabov, M.~Madigan, L.~Mantani, M.~Morales Alvarado, J.~M.~Moore and
M.~Ubiali (PBSP), Eur.\ Phys.\ J.\ C {\bf 84} (2024) 805, arXiv:2402.03308 [hep-ph].

\bibitem{iranipour}
S.~Iranipour and M.~Ubiali, \Journal{JHEP}{05}{032}{2022}, arXiv:2201.07240 [hep-ph].

\bibitem{nnpdf40}
R.~D.~Ball {\it et al.} (NNPDF), Eur.\ Phys.\ J.\ C {\bf 82} (2022) 428, arXiv:2109.02653 [hep-ph].

\bibitem{eicfpf}
E.~Hammou and M.~Ubiali, Phys.\ Rev.\ D {\bf 111} (2025) 095028, arXiv:2410.00963 [hep-ph].

\end{thebibliography}
\end{document}